\documentclass[%
 reprint,
 amsmath,amssymb,
pra,
]{revtex4-1}
\usepackage{graphicx}% Include figure files
\usepackage{bm}% bold math
\usepackage{esvect}
\usepackage{xcolor}
\usepackage{amsmath}
\usepackage[colorlinks=true, citecolor=blue, linkcolor=blue, urlcolor=blue]{hyperref}
\usepackage{cleveref}
\begin{document}

%\preprint{APS/123-QED}

%\title{Coherent Kapitza-Dirac effect for proton beam}% Force line breaks with \\
\title{Time evolution of charged particle wave functions in optical crystal: The coherent Kapitza-Dirac effect for plasma-based proton beams}

\author{Sushanta Barman}
%\email{sushanta@iitk.ac.in}
\author{Sudeep Bhattacharjee}%
 \email{sudeepb@iitk.ac.in}
\affiliation{%
 Department of Physics, Indian Institute of Technology - Kanpur, Kanpur 208016, India}%


%\affiliation{
%\affiliation{
%\affiliation{%



\begin{abstract}
The stationary eigenstates and eigenvalues for the ponderomotive potential of an optical crystal confined in a one-dimensional infinite square well are numerically obtained. The initial states of the incoming particles taken as Gaussian, are expanded in the basis of the stationary eigenstates of the ponderomotive potential, to obtain the subsequent time evolution of the wave function of the particle during the interaction with the optical crystal. From the results of the time evolution of the probability density, it is observed that the particles get localized at equidistant positions in the transverse direction, which results in the diffraction pattern. The temporal evolution of the diffraction pattern is analyzed. As an application, the diffraction of proton beams is studied, where the experimental parameters are optimized to observe the diffraction pattern for a microwave plasma-based proton beam system. The observations are important for design of proton based matter-wave interferometers.
%\begin{description}
%You may use the \texttt{description} environment to %structure your abstract;
%use the optional argument of the \verb+\item+ command to %give the category of each item. 
%\end{description}
\end{abstract}

\maketitle

%\tableofcontents
%================================================================
% Introduction
%================================================================
\section{\label{sec:introduction}Introduction}
The wave-particle duality of matter is a central part of quantum mechanics. de Broglie, in 1924 \cite{Broglie,Weinberger,Edward}, proposed that matter can exhibit wave-like behavior. Shortly afterward, Davisson and Germer (1927) observed electron diffraction from the periodic structure of crystal lattice \cite{Davisson}, which was the direct evidence of the wave-particle duality of electrons. After these observations, a few years later, Kapitza and Dirac (1933) \cite{Kapitza} proposed that the regular structures of electromagnetic standing waves can be used as a grating to diffract electrons. This can be explained in terms of the wave-particle duality of the electrons. The matter waves of electrons are diffracted by the optical crystal, where the regions of maximum intensity of the standing waves act like the crystal planes.  The estimated strength of the diffracted beam over the straight beam, using the standing wave of green mercury light was 10\textsuperscript{-14} \cite{Kapitza}. Therefore, to observe detectable diffracted beams, light sources of higher intensities are needed. For this reason, attempts for the experimental observation of the Kapitza-Dirac (KD) effect had to wait for the discovery of the laser light. Efforts were made after the development of the laser to observe the KD effect \cite{Schwartz1,Bartell,Takeda,Pfeiffer, Schwartz2}. In most of the attempts, the diffraction patterns could not be resolved due to lack of proper optimization of the experimental parameters \cite{Schwarz3,Fedorov}.\par

D. E. Pritchard at MIT \cite{Philip1, Philip2, David, Gould, Peter1, Peter2} in 1986 could experimentally demonstrate the diffraction of Na atoms by a near-resonant standing waves of laser light. Using a supersonic atomic beam with extremely small transverse momentum, diffraction up to 10 $\hbar$k was resolved. Finally, it took more than 60 years after the first proposal, the diffraction of electron beams by electromagnetic standing waves was observed by H. Batelaan in 2001 \cite{Freimund8,Freimund9}. It is also realized that the diffracted beams are coherent to each other \cite{Freimund8}. Hence, this type of set-up is extremely important for interferometric systems \cite{Ernst,Giltner,Berman,Alexander}. Atoms can be diffracted by material gratings \cite{Perreault} such as those for neutrons \citep{Mason}, as well as the phase gratings formed by the standing wave of electromagnetic radiations \cite{Philip1,Philip2}. However, for the diffraction of charged particles, optical crystals of the standing wave of light is a natural choice, because small electric and magnetic fields due to the presence of materials can smear out the diffraction peaks \cite{H_Batelaan2}.\par 

KD effect continues to be a subject of active research and finds wide applications in areas such as matter-wave interferometers \cite{McDonald}, general relativity \cite{Savas1,Renee,Altin,Qingqing}, gravitational waves \cite{Savas2,Hogan,Colella,Tarallo,Muntinga,Nesvizhevsky,Peter_Asenbaum}, determination of fine structure constant \cite{Malo,Rym} including non-perturbative quantum dynamical systems \cite{Rym}. It has been demonstrated recently that the KD effect on optical standing waves formed by the fundamental frequency and it's third harmonic can be used as an electron beam splitter \cite{Asma}. Spin flips in the Compton scattering during the interaction with optical crystal has been studied by D. Seipt et al.(2018) \cite{Seipt} using the density matrix formalism. The space charge effect on the interaction of the charged particles with a laser has been demonstrated in Ref. \cite{Dabagov}. Theoretical feasibility of the spin-dependent KD effect for different polarization geometries of the counter-propagating laser beams has been studied \cite{Ahrens,Matthias}. \par  

Attempts were made earlier to explain the KD effect by the wave theory \cite{H_Batelaan}, where the states of the charged particles are expanded in the basis of plane waves, quantum and classical description of electron motion \cite{Efremov}. The scattering of electron by traveling waves has also been studied using Helmholtz–Kirchhoff theory of diffraction \cite{Armen}. However, there is a lack of quantum mechanical study of the ponderomotive potential and in particular the time evolution of the matter wave function by utilizing the actual eigenvalues and eigenstates of the ponderomotive potential. In order to have a better understanding of the dynamics of the charged particles in the fields of the optical crystal, it is necessary to investigate the eigenstates and eigenvalues of the ponderomotive potential, which can be helpful to understand the mechanism of the diffraction phenomena. A knowledge of the stationary eigenstates of the ponderomotive potential can be used as a complete basis to expand the state of the charged particles in optical crystal, and calculate the time evolution of the state. As an application, the diffraction of proton beams (heavier mass and opposite polarity than the electrons) by the standing wave of a Nd:YAG laser, which has not been studied earlier is investigated. It will be interesting to investigate the possibilities of observing the KD effect for the plasma-based proton beams, which can be helpful in designing proton based interferometers. \par 

In this article, we have presented the eigenstates and eigenvalues of the ponderomotive potential of the standing waves of laser, confined in an one-dimensional infinite square well, by solving the time-independent Schrodinger's equation. To estimate the position of the diffraction orders and peak amplitudes, the initial states of the charged particles are taken as Gaussian and expanded in the basis of the stationary eigenstates of the pondermotive potential. Time evolution of the charged particle wave function in the optical crystal is obtained by solving the time-dependent Schrodinger's equation. The results of the time evolution of probability amplitude indicate that during the interaction, the charged particles get localized at equidistant positions in the transverse direction resulting in the diffraction pattern. To observe the KD effect for protons, the experimental parameters of optical crystal and proton beams are optimized, which will be helpful to design the experimental set-up to investigate the KD effect for plasma-based proton beams and spin dynamics of charged particles in the optical crystal. The design will complement the existing plasma ion source in the laboratory \cite{Mathew,sanjeev2}. \par

This article is organized as follows. In Sec. \ref{subsec:the_ponderomotive_potential}, an expression for the pondermotive potential is derived in terms of the experimental parameters to work with Schrodinger's equation. In Sec. \ref{sec:Eigenstates_&_Eigenvalues}, the theoretical background is discussed and simulation techniques to calculate the eigenstates and eigenvalues for the ponderomotive potential are introduced. In Sec. \ref{sec:time_evolution}, the time evolution of a charged particle wave function during the interaction with optical crystal is studied, which explains the KD effect. The possibilities to observe the diffraction of the plasma-based proton beams are studied in Sec. \ref{sec:Parametric_study}. The experimental parameters are optimized for the diffraction of proton beams. Finally, in Sec. \ref{sec:conclusions}, the results are summarized and important  concluding remarks are provided.

\begin{figure}[h]
    \centering
    \includegraphics[width=8.6cm]{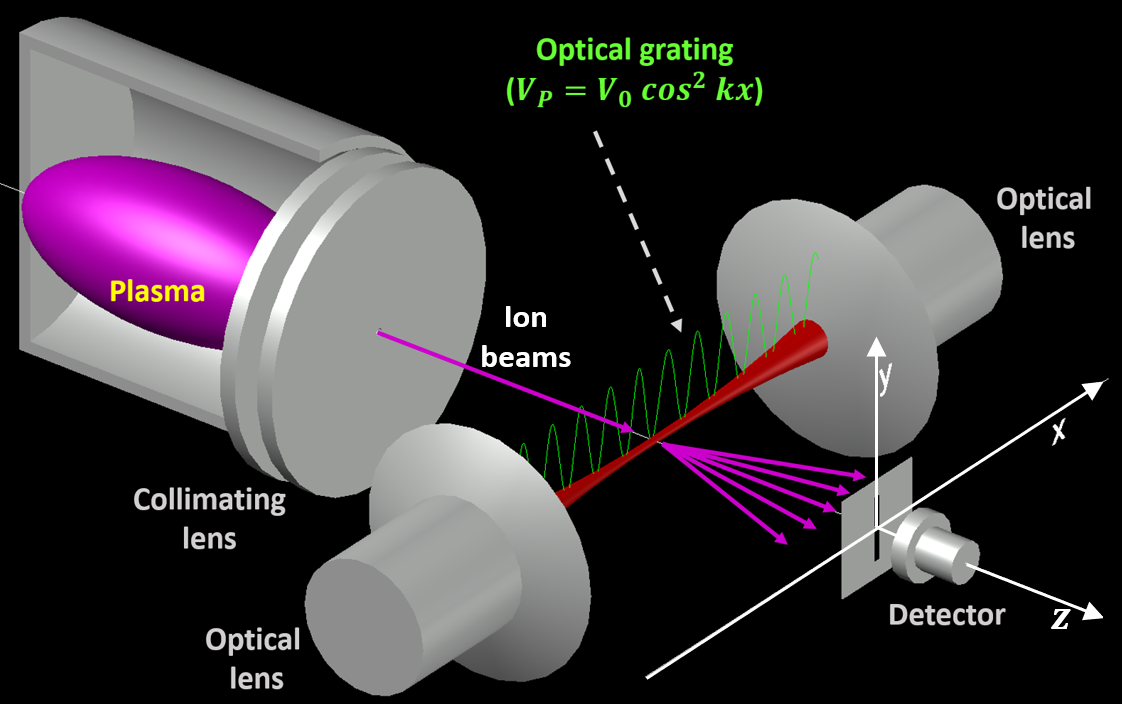}
    \caption{The schematic diagram of the experimental set-up for the diffraction of plasma ion beams by standing wave of a laser. The counter-propagating ($x$ direction) laser waves are focused to form a standing wave with the help of optical lenses. The incident charged particles interact with the optical crystal, and the diffracted particles can be detected by the detector at different positions along the $x$ axis to resolve the diffraction pattern. The beams, laser interaction region, the diffracted beams, and the detector are all kept in vacuum.}
    \label{fig:kd_setup}
\end{figure}

%===============================================================
% The Ponderomotive potential
%===============================================================
\section{\label{subsec:the_ponderomotive_potential} The Ponderomotive potential}
Charged particles placed in an inhomogeneous electromagnetic field experience a nonlinear force known as ponderomotive force which displaces the particle towards the region of weaker field strength \cite{H_Batelaan,Chan}. The displacement of the particle occurs due to the unequal magnitude of the force during both half periods at different positions. The magnitude of the force is higher in the stronger field region, and it does not cancel out in both the half-cycle \cite{H_Batelaan}. Due to this ponderomotive force, the particles gain a transverse momentum, which results in the deflection of the particles along the transverse direction. To work with the Schrodinger's equation, we need to derive an expression for ponderomotive potential energy in terms of the experimental parameters. \par

The expression for ponderomotive potential energy can be derived by considering the harmonic motion of the charged particles in the electromagnetic fields and calculating the average energy of the particle over a full period of the time. \par

A schematic diagram of the experimental set-up is shown in Fig. \ref{fig:kd_setup}, where the charged particles, extracted from a microwave plasma source, are focused by an electrostatic lens. The particles moving along $z$ direction interact with a standing wave of light formed by two counter-propagating laser parallel to $x$ axis. To achieve higher intensities at the interaction region, the laser beams are focused by two optical lenses. The vector potential of the standing wave can be expressed as \cite{H_Batelaan}
\begin{equation}
A_z=A_{0} \cos(kx) \sin(\omega t),
\end{equation}
where $k$ and $\omega$ are the wave vector and angular frequency of laser light respectively, and $A_0$ is the amplitude of vector potential which can be calculated from the laser intensity ($I=\epsilon_0 c \omega^2 A^2_0 /2$).
The electric and magnetic fields can be obtained as
\begin{align} 
E_z=-\frac{\partial A_z}{\partial t}=- A_0 \omega \cos(kx) \cos(\omega t),\nonumber\\
B_y=-\frac{\partial A_z}{\partial x}= A_0 k \sin(kx) \sin(\omega t).
\end{align}
The resultant electric and magnetic fields are $\pi/2$ out of phase in both space and time.
\par
%@@@@@@
\begin{figure}
\centering{
\includegraphics[width=8.6cm]{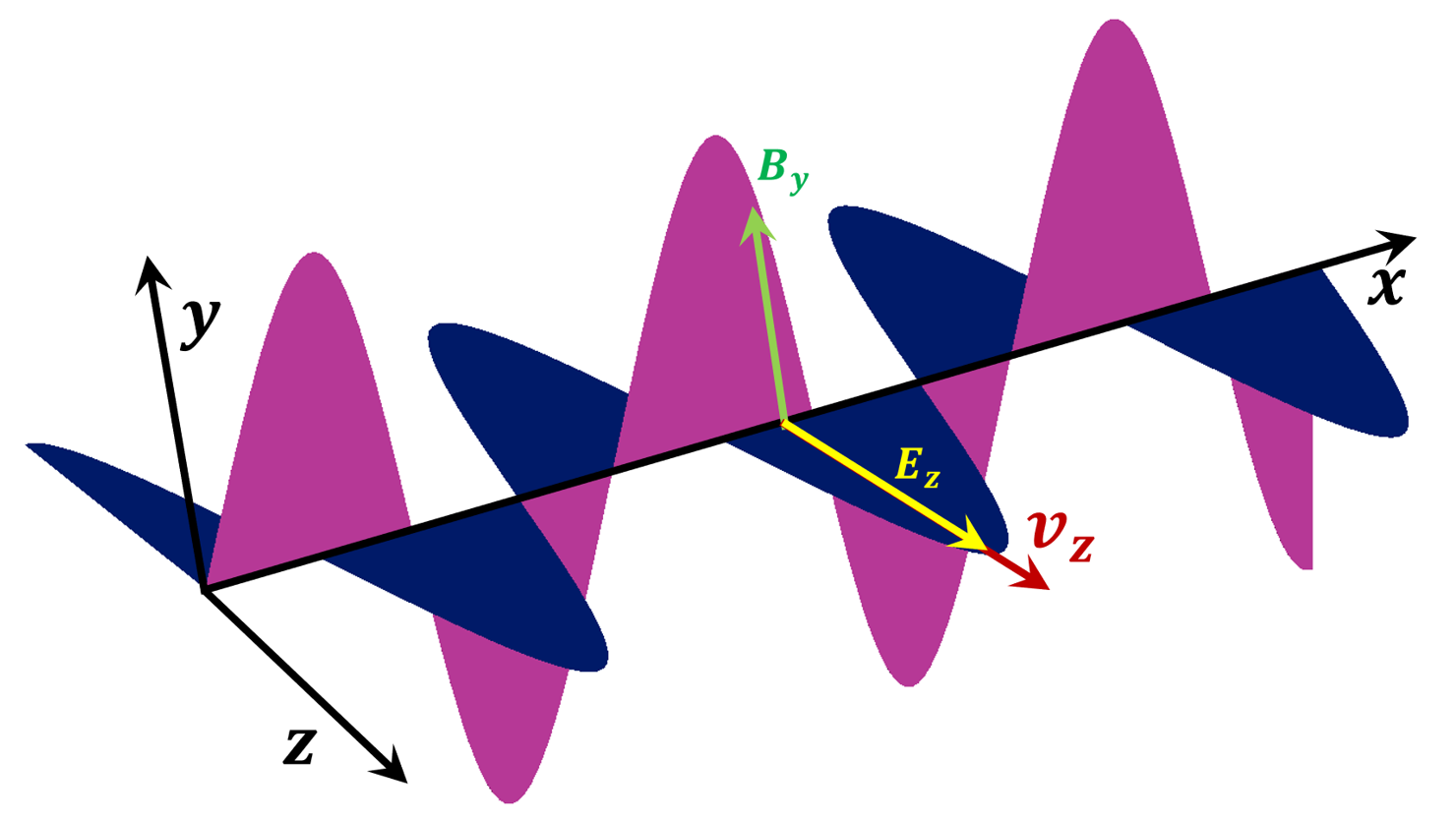}
}
\caption{The electric field ($E_z$) and magnetic field ($B_y$) of the standing wave. The charged particle oscillates along the $z$ axis due to $E_z$, and it experiences a pondermotive force along $x$ direction. This force deflects the particles in the transverse direction ($x$ axis).}
\label{fig:em_fields}
\end{figure}
%@@@@@@
Charged particles in the standing wave will oscillate along the $z$ direction due to the oscillating electric field which is shown in Fig. \ref{fig:em_fields}, and the equation of motion along $z$ is given by
\begin{equation}
m\ddot{z}=-eE_z,
\label{eq:equation_of_motion}
\end{equation}
where $m$ is the mass of electron. The solution of Eq. (\ref{eq:equation_of_motion}) is found to be
\begin{equation}
z=-\frac{eA_0}{m\omega} \cos(kx) \cos(\omega t).
\end{equation}
The time averaged mean-squared displacement of the particles can be obtained as
\begin{equation}
\langle z^2 \rangle=\frac{1}{2} \left(\frac{eA_0}{m \omega} \right)^2 \cos^2 (kx).
\end{equation}
Since the particle is in harmonic motion, time averaged potential energy of the particles is given by
\begin{equation}
V_p=\frac{1}{2} m \omega^2 \langle z^2 \rangle = \frac{e^2 A^2_0}{4m} \cos^2 (kx).
\label{eq:averaged_potential}
\end{equation}
Using the expression of laser intensity, $I=\epsilon_0 c \omega^2 A^2_0 /2$, Eq. (\ref{eq:averaged_potential})  can be expressed as
\begin{equation}
V_p=\frac{e^2 I}{2 m \epsilon_0 c \omega^2} \cos ^2 (kx)=V_0 \cos^2 (kx),
\label{eq:V_p}
\end{equation}
where $V_0=e^2 I / 2m \epsilon_0 c \omega^2$ is the strength of the ponderomotive potential which depends on $I$, $m$ and $\omega$; it is independent of the polarity of the charged particles.\par

To calculate the time evolution of the charged particle wave functions in the optical crystal, the ponderomotive potential of Eq. (\ref{eq:V_p}) can be employed in the time dependent Schrodinger's equation.
%@@@@@@
\begin{figure}[h]
    \centering
    \includegraphics[width=8.6cm]{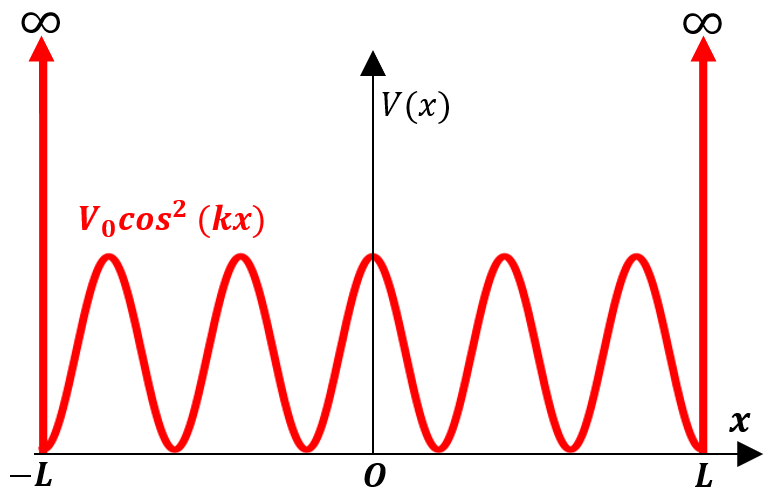}
    \caption{Variation of the effective potential $V(x)$ with $x$. The ponderomotive potential $V_P (x)$ is confined in an infinite square well having width 2$L$.}
    \label{fig:ponderomotive_potential}
\end{figure}

%===============================================================
%Eigenstates and eigenvalues of the ponderomotive potential
%===============================================================
\section{\label{sec:Eigenstates_&_Eigenvalues}Eigenstates and eigenvalues of the ponderomotive potential}
To estimate the position of different orders and peak amplitudes in the diffraction pattern, it is necessary to understand the particle dynamics using quantum mechanical treatments. Before studying the particle dynamics, we investigate the stationary eigenstates and eigenvalues of the ponderomotive potential, which is important for understanding the diffraction phenomena. The diffraction phenomena can be explained in terms of the eigenstates of the ponderomotive potential. \par

%###########################
The divergence of the incoming particle beams due to space charge repulsion force can smear out the diffraction pattern. Hence, it is necessary to look at the contribution of the repulsion force on the particle dynamics. The expression for the magnitude of the  ponderomotive force can be obtained as
\begin{equation}
F_P = - \frac{dV_p}{dx} = V_0 k \sin{\left(2kx \right)}.
\label{eq:ponderomotive_force}
\end{equation}
The maximum value of $F_P$ comes out to be $V_0 k\sim 10^{-14}$ N for the standing wave of laser having wavelength $\lambda = 532$ nm, power = 0.2 J/pulse, beam waist at the interaction region = 125 $\mu$m, and pulse width = 10 ns. \par

The number of particles present in the laser fields during the interaction time ($\tau=0.8$ ps) depends upon the beam current. Typically for a beam current of 100 nA the number is calculated to be 1. The coulomb repulsion between two charges in the interaction region of the focused beam can be calculated by
\begin{equation}
\left|F_c \right|=\frac{1}{4 \pi \epsilon_0} \frac{e^2}{s^2},
\label{eq:coulomb_force}
\end{equation}
where $s$ ($\sim 2$ $\mu$m) is the typical diameter of the incident particle beams. The magnitude of $F_c$ comes out to be $\sim 10^{-17}$ N. The relative magnitude of the two forces is calculated as $\left| F_P / F_c \right| \sim 10^{3} $.  Therefore, for such lower beam currents ($< 100$ nA), the space charge effects can be neglected. Hence, in rest of the calculations, the effect of the space charge repulsion force is neglected. \par
%#######################
\begin{figure}[b]
    \centering
    \includegraphics[width=8.6cm]{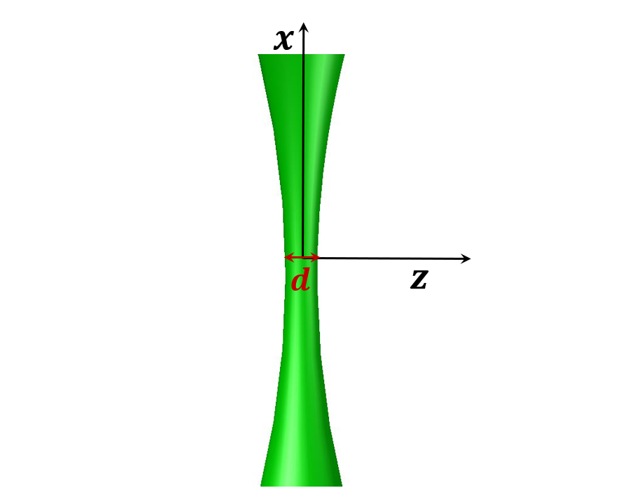}
    \caption{Beam waist ($d$) of the standing wave of laser at the interaction region. In the numerical calculation, $d$ is taken as 125 $\mu$m.}
    \label{fig:laser_beam_waist}
\end{figure}
%@@@@@@
\begin{figure}[h]
    \centering
    \includegraphics[width=8.6cm]{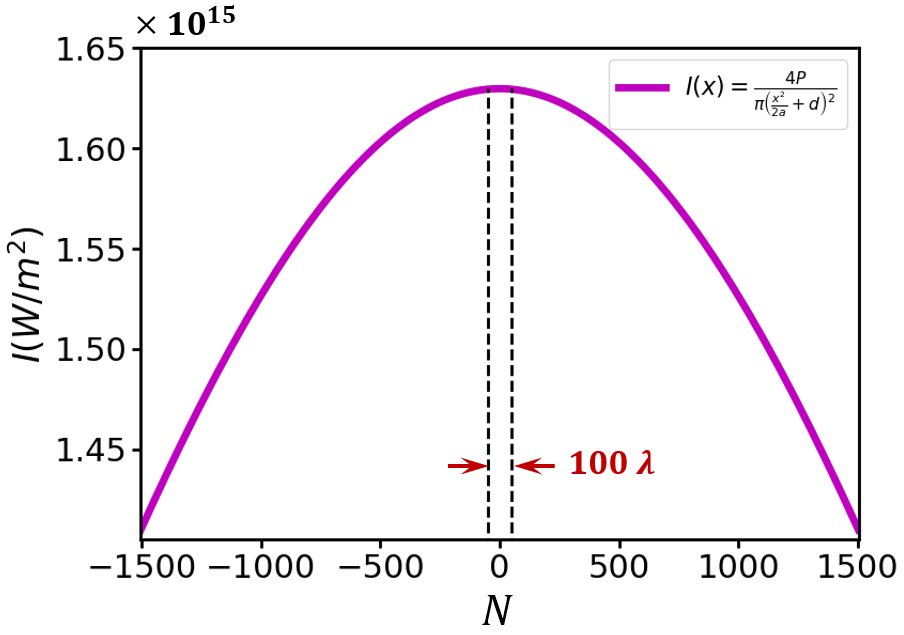}
    \caption{Variation of the intensity of focused laser beam along the transverse direction ($x$). $N$ is the number of wavelengths $\lambda$'s along the $x$ direction (Negative sign indicates other side of the peak). Here, the curvature of the focused laser beams is considered to be parabolic, $x^2 = 4a(z-d/2)$ where $a$ ($=3.4 \times 10^{-3}$) is calculated using the known values of the laser beam sizes before focusing by a converging lens (6 mm) and at the interaction region ($d$).}
    \label{fig:intensity_profile}
\end{figure}
\begin{figure*}
    \centering
      \includegraphics[width=17.2cm]{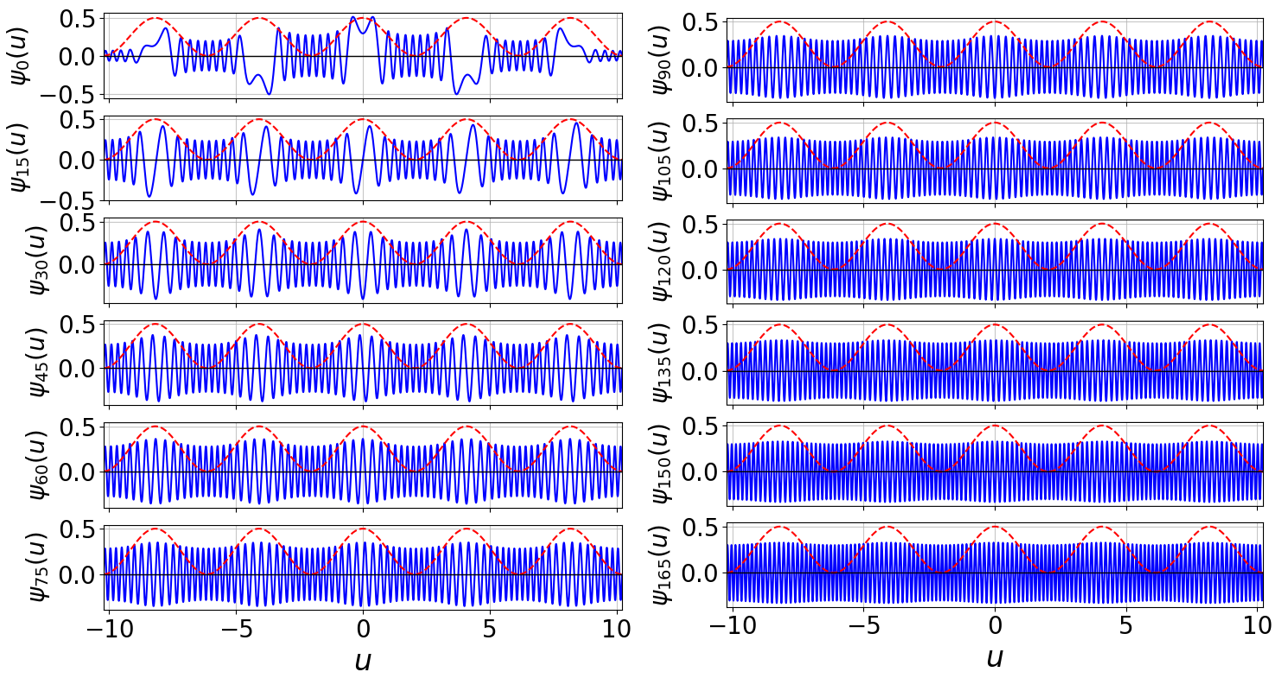}
    \caption{The normalized wavefunctions for $n$= 0, 15, 30, 45, 60, 75, 90, 105, 120, 135, 150, and 165. It is clear that for even values of $n$, the wave functions are symmetric and for odd values of $n$, the wave functions are anti-symmetric. Also, the wave functions become highly oscillatory for higher values of $n$, because the wave vector ($K$) increases with $n$ which is shown in Fig. \ref{fig:eigenvalues}.}
    \label{fig:wavefunctions}
\end{figure*}

%===============================================================
%Theory and Numerical Methods
%===============================================================
\subsection{\label{subsec:theory_simulation}Theory and Numerical Methods}
The one dimensional time independent Schrodinger's equation for the particles in the ponderomotive potential is given by
\begin{equation}
\begin{aligned}
-\frac{\hbar^2}{2m}\frac{d^2\psi}{dx^2} + V_0 \cos^2(kx) \psi= E \psi.
\label{eq:time_independent_Sch_equn}
\end{aligned}
\end{equation}

The exact analytical solution of Eq. (\ref{eq:time_independent_Sch_equn}) is difficult, however using numerical computation, the equation can be solved under certain conditions and after scaling the problem. The system under consideration is as shown in Fig. \ref{fig:ponderomotive_potential}, in which the ponderomotive potential is confined in a one dimensional infinite square well so that all the bound states and corresponding energy eigenvalues of the system can be calculated numerically by solving the time independent Schrodinger's equation. Therefore, the effective potential becomes
\begin{align}
V(x) &= \begin{cases}
            V_0 \cos^2 (kx)  \text{ for $-L < x <L$}\\
            \infty \text{ otherwise},
        \end{cases}
\end{align}
where $L$ is the half width of the infinite square well. We are interested to solve Eq. (\ref{eq:time_independent_Sch_equn}) in the region $-L < x < L$. $V_0$ is taken to be constant within this range, because of the parabolic curvature ($x^2 = 4a(z-d/2)$) of the focused laser beams as shown in Fig. \ref{fig:laser_beam_waist}, where $a \sim 3.4 \times 10^{-3}$. The intensity of the laser beams can be obtained as,
\begin{equation}
I(x)=\frac{4P}{\pi \left( \frac{x^2}{2a} +d   \right)^2}.
\label{eq:laser_intensity}
\end{equation}
It is clear that the intensity of the beam is almost constant in the interaction region as shown in Fig. \ref{fig:intensity_profile} ($2L \sim 2.5 \lambda$). The strength of the potential $V_0$ is calculated for the standing waves of the laser having power ($P$) = 0.2 J/pulse, pulse width = 10 ns, wavelength = 532 nm, and beam waist ($d$) at the interaction region = 125 $\mu$m (Fig. \ref{fig:laser_beam_waist}). Intensity ($I \sim 4P/\pi d^2$ for $x^2/2a << 1$) of the laser beam in the interaction region is calculated to be $\sim$1.6 $\times$ 10 \textsuperscript{15} W/m\textsuperscript{2}. The calculated value of $V_0$ comes out to be $\sim 4.29 \times 10^{-3}$ eV, which is quite small. Hence,  it is necessary to scale the problem to increase the resolution in the numerical calculation. \par 

In order to scale the problem, we define three new parameters, $\beta=2mE/\hbar^2$, $\alpha^2=2mV_0/\hbar^2A$ ($A$ is a constant), and $E'=\beta/\alpha^2	$. A new variable is defined as $u=\alpha x$. After substitution of the parameters and changing the variable $x$, Eq. (\ref{eq:time_independent_Sch_equn}) becomes
\begin{equation}
\label{eq:normalised_1}
\frac{d^2\psi}{du^2}+\left\lbrace E'- A \cos^2\left(\frac{ku}{\alpha} \right)  \right\rbrace \psi=0.
\end{equation}

Thereafter, substituting the values of the wave vector $k$ and $\alpha$ in Eq. (\ref{eq:normalised_1}), we obtain
%@@@@@@@@@@
\begin{equation}
\frac{d^2 \psi}{du^2} + \left\lbrace E' - A \cos^2 \left( 3.433 \times 10^{-2} \times \sqrt{A} u \right) \right\rbrace \psi = 0.
\label{eq:mormalized_2}
\end{equation}
%@@@@@@
\par 

The above second order equation for highly oscillatory potential can be solved numerically. At first, Eq. (\ref{eq:mormalized_2}) is written as two coupled first order equations,
%@@@@@@@@@@
\begin{equation}
\frac{d\psi(u)}{du} = \phi (u)  \text{ and } \frac{d\phi (u)}{du}=f(u),
\label{eq:coupled_equation}
\end{equation}
%@@@@@@
where $f(u)=- \left( E' - A \cos^2 ( 3.433 \times 10^{-2} \times \sqrt{A} u )\right)$. \par

The boundary conditions for $\psi (u)$ in Eq. (\ref{eq:coupled_equation}) are $\psi (-L) = \psi(L)=0$. But this condition does not guarantee the continuity of $\phi(u)$. Hence, in the numerical calculation using Runge-Kutta (RK4) method, a different set of boundary conditions for $\phi (u)$ and $\psi (u)$ have been chosen. The effective potential $V(u)$ is symmetric with respect to the origin ($u=0$). Therefore, it is expected that even wave functions should be symmetric and odd wave functions should be antisymmetric with respect to the same point \cite{Griffiths}. Hence, we have chosen the following boundary conditions, 
\begin{subequations}
\begin{align}
\phi_n\left(u=0\right) & = 0 \text{  for even $n$} \label{eq:BC1}, \\
\psi_n \left(u=0 \right) & = 0 \text{ for odd $n$}\label{eq:BC2}.
\end{align}
\end{subequations}

In the numerical calculation, the half-width of the infinite well $L$ is taken as 10 units and $A=500$. Any value of $A$ can be chosen for convenience in the calculation, because the actual eigenvalues ($E$) are given by $E=V_0 (E'/A)$, therefore the change in $A$ will be self-adjusted by the eigenvalues $E'$ of Eq. (\ref{eq:mormalized_2}) and the actual eigenvalues ($E$) of the system will remain the same. \par

\begin{figure}[h]
    \centering
    \includegraphics[width=8.6cm]{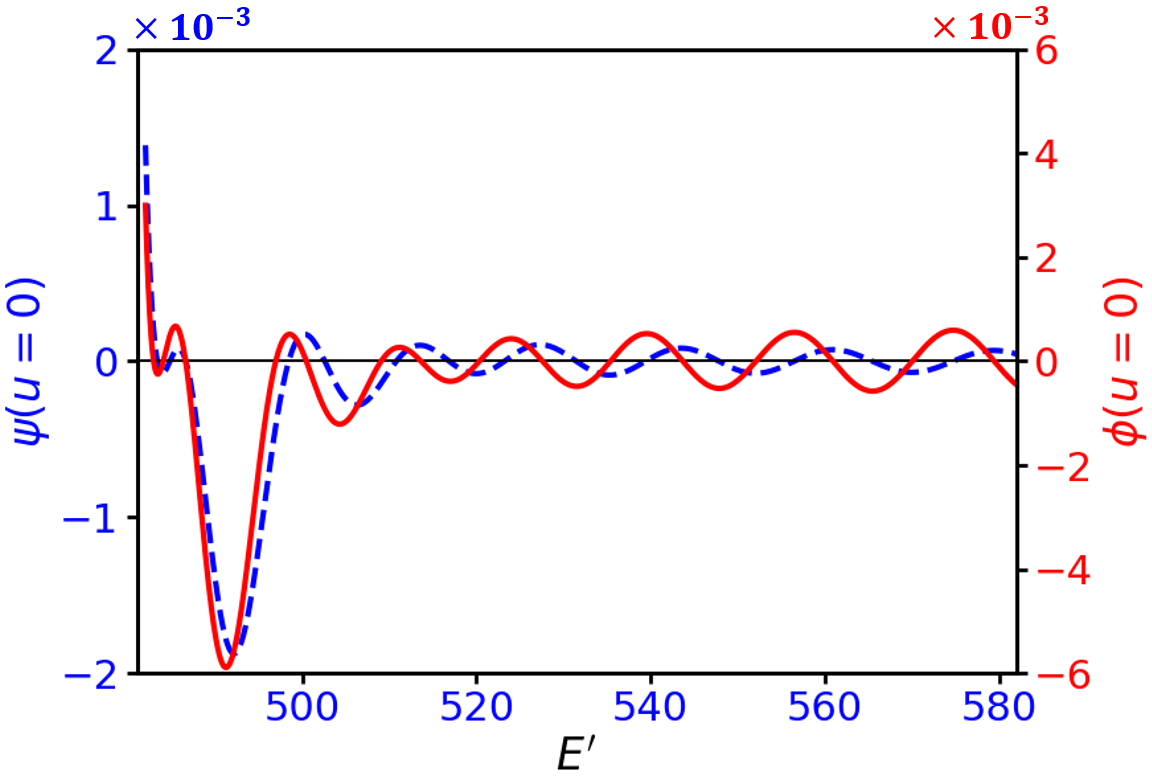}
    \caption{The solution $\psi (u)$ (blue-dotted line) and $\phi (u)$ (solid red line) at the mid point ($u=0$) of the coupled Eq. (\ref{eq:coupled_equation}) for different values of the scaled energy $E'$. The points, where $\psi (u=0)$ changes sign corresponds to the odd eigenvalues. Similarly, the change in sign of $\phi (u=0)$ gives the even eigenvalues.}
    \label{fig:energy_scan}
\end{figure}

\begin{figure}[h]
    \centering
    \includegraphics[width=8.6cm]{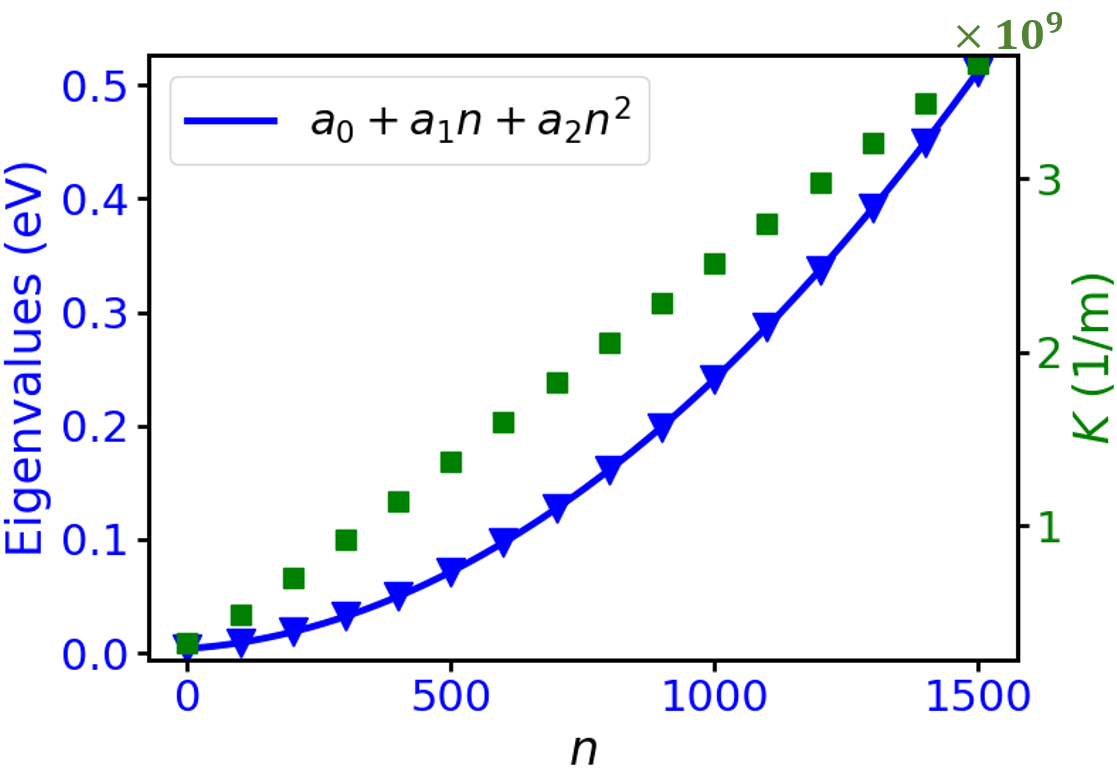}
    \caption{Variation of eigenvalues $E$ (blue-colored triangles) for ponderomotive potential with $n$ as well as of the wave vector $K$ (green-colored squares) of the corresponding eigenstates with $n$. Blue-colored line represents the second order polynomial fit of the eigenvalues with the fitting parameters: $a_0 =3.75 \times 10^{-3}$ eV, $a_1$ = 3.415 $\times$ 10\textsuperscript{-5} eV, and $a_2$ = 2.0297 $\times$ 10\textsuperscript{-7} eV.}
    \label{fig:eigenvalues}
\end{figure}

To search for the eigenvalues for the ponderomotive potential, we have introduced a highly accurate method in which, at first, the approximate eigenvalues are found so that the boundary conditions (\ref{eq:BC1}) and (\ref{eq:BC2}) are satisfied. The variations of $\psi(u=0)$ and $\phi(u=0)$ with $E'$ are shown in Fig. \ref{fig:energy_scan}. The values of $E'$ for which $\psi(u=0) = 0$, gives the odd eigenvalues and similarly $\phi(u=0) = 0$, corresponds to the even eigenvalues. From Fig. \ref{fig:energy_scan}, it is clear that the even and odd eigenvalues occur alternately. Thereafter in the next step, the eigenvalues are recalculated near to the eigenvalues obtained in the first step by decreasing the step size ($\Delta E'$) by an order. In the subsequent steps, this procedure is repeated by lowering the step size to obtain the eigenvalues with resolution $\Delta E' = 10^{-10}$. The calculated wave functions and corresponding energy eigenvalues are shown in Fig. \ref{fig:wavefunctions} and Fig. \ref{fig:eigenvalues}, respectively.

%@@@@@@
\begin{figure}
    \centering
    \includegraphics[width=8.6cm]{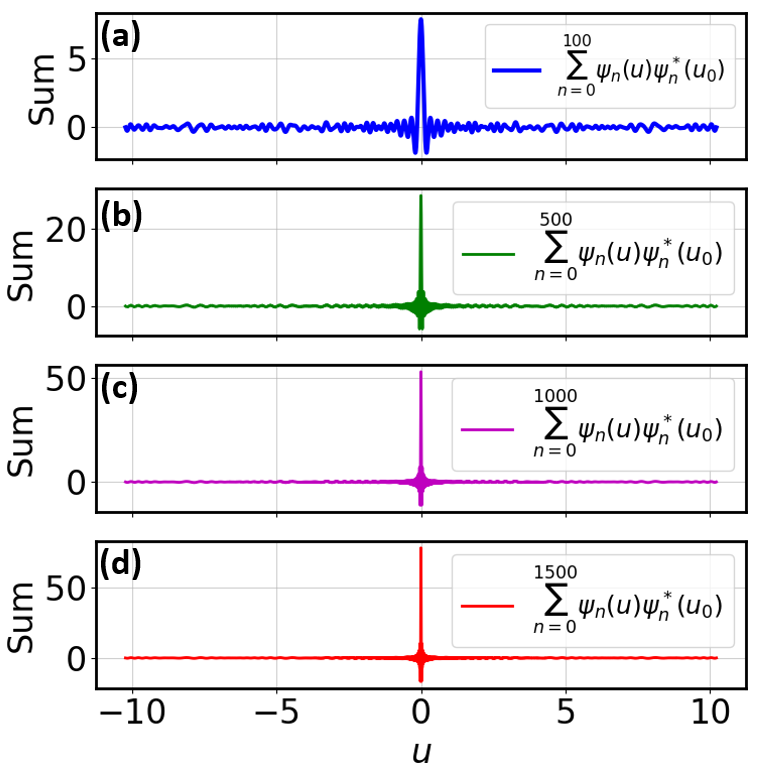}
    \caption{Completeness property of the stationary states for (a) $n=100$, (b) $n=500$, (c) $n=1000$, and (d) $n=1500$. As the number of wave functions increases, the summation in the left hand side of Eq. (\ref{eq:completeness}) becomes like a delta function.}
    \label{fig:completeness}
\end{figure}
%@@@@@@
%===============================================================
% Results and Discussions
%===============================================================
\subsection{\label{subsec:results}Results and Discussions}
\subsubsection{\label{subsec:eigenstates}Eigenstates}
We have calculated 1500 stationary states for the ponderomotive potential and some of the normalized states are shown in Fig. \ref{fig:wavefunctions}. It is observed that the eigenstates are symmetric for even values of $n$ and antisymmetric for odd values of $n$, which is expected because the potential is symmetric with respect to the origin. It is also observed that as $n$ increases the wave functions become highly oscillatory.
\par
It will be interesting to check the completeness property of the stationary eigenstates of the pondermotive potential. The completeness relation of a set of wave functions is given by \cite{Zwiebach}
\begin{equation}
\sum_{n} \psi_n(u) \psi^*_n(u_0)=\delta (u-u_0),
\label{eq:completeness}
\end{equation}
where $u_0$ is any arbitrary point in the region of interest. In the numerical calculation, $u_0$ is chosen to be 0. To check the completeness property of the calculated wave functions, the left hand side of Eq. (\ref{eq:completeness}) is calculated using 1500 normalized wave functions and is shown in Fig. \ref{fig:completeness}. From figure \ref{fig:completeness}, it is clear that with increase in the number of eigenstates included in the summation, the left hand side of Eq. (\ref{eq:completeness}) becomes closer to a delta function, which tells about the correctness of the calculation. These wave functions can be used as a basis set to expand the state of the charged particle during the interaction with stationary fields of the optical crystal.
\subsubsection{\label{subsec:eigenvalues}Eigenvalues}
The eigenvalues of the stationary states for the ponderomotive potential is calculated, and shown in Fig. \ref{fig:eigenvalues}. It is found that the eigenvalues are fitted well with the polynomial $E=a_0 + a_1 n +a_2 n^2$, where $n=$0, 1, 2... up to 1500. The values of the fitted parameters are $a_0 =3.75 \times 10^{-3}$ eV, $a_1$ = 3.415 $\times$ 10\textsuperscript{-5} eV, and $a_2$ = 2.0297 $\times$ 10\textsuperscript{-7} eV. In the case of a particle in an infinite square well, the eigenvalues vary as $E_n=n^2 \pi^2 \hbar^2/8mL^2$, where $L$ is the half-width of the well. For particle in a harmonic well, the eigenvalues vary as $E_n = (n+\frac{1}{2})\hbar \omega$. From the fitting, it is observed that the eigenvalues for the ponderomotive potential confined in an infinite square well vary as a quadric polynomial of $n$, which is a mixture of both the infinite well and the harmonic oscillator. Because of the confinement of the ponderomotive potential by the infinite square well, the infinite well also contributes to the eigenvalues of the system. Their contributions are determined by the coefficients $a_1$ and $a_2$. From the values of the fitted parameters, it is found that the contribution of the infinite well is $\sim$0.59$\%$ of the contribution of the ponderomotive potential, and it can be reduced by increasing the width ($2L$) of the infinite square well in the simulation. The wave vector $K$ of the eigenstates for the ponderomotive potential increases with $n$, as shown in Fig. \ref{fig:eigenvalues}.
%===============================================================
% Time evolution of the wave functions
%===============================================================
\section{\label{sec:time_evolution}Time evolution of the wave functions}
In order to predict the position of the diffraction orders and peak amplitudes, it is necessary to observe the charged particle dynamics in the nonlinear fields of the standing waves of the laser. The time evolution of the initial state of the incoming charged particles are calculated by solving the time dependent Schrodinger's equation. \par

The one dimensional time dependent Schrodinger's equation for the charged particle is given by
\begin{equation}
\left(-\frac{\hbar^2}{2m} \frac{\partial^2}{\partial u^2} + V_0 \cos^2 ku \right) \psi_e (u,t)= i \hbar \frac{\partial \psi_e (u,t)}{\partial t}, 
\label{eq:time_dep_sch}
\end{equation}
where $\psi_e(u,t)$ is the wave function of the charged particles. \par 

The stationary eigenstates of the ponderomotive potential is used as a complete set of basis to expand the states of the charged particles in the fields of the optical crystals, and the solution of Eq. (\ref{eq:time_dep_sch}) can be expressed as \cite{Griffiths}
\begin{equation}
\psi_e(u,t)=\sum_{n=0}^{1500} c_n e^{-\frac{i E_n t}{\hbar} } \psi_n (u),
\label{eq:expansion}
\end{equation}
where $\psi_n (u)$ and $E_n$ are the $n$th stationary eigenstate and eigenvalue of the ponderomotive potential confined in an infinite square well, respectively. $c_n$'s are the coefficients which can be calculated from the initial conditions.
\par 

In the simulation we have taken the initial state of the particle as Gaussian, 
\begin{equation}
\psi_e(u,0)=\frac{1}{\sqrt{2 \pi \sigma}} e^{\frac{(u-u_0)^2}{2 \sigma^2}},
\end{equation}
where $\sigma$ = 0.07 and $u_0$ = 0.
\par 
%@@@@@@@@@@@@@@@@@@@@
\begin{figure}
    \centering
    \includegraphics[width=8.6cm]{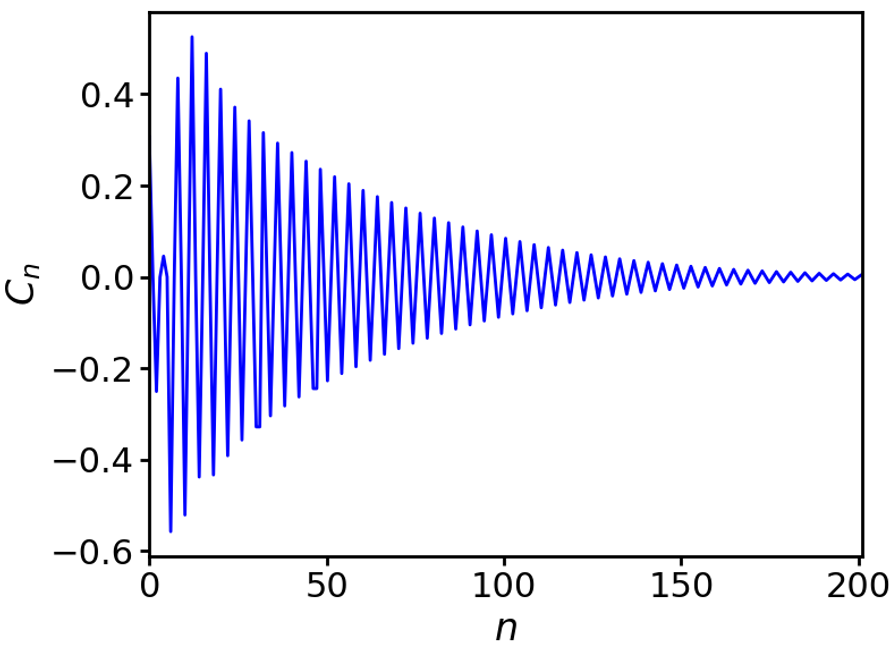}
    \caption{Variation of $c_n$'s with $n$.}
    \label{fig:cns}
\end{figure}
%@@@@@@@@@@@@@@@@@@@@@
The expansion coefficients in Eq. (\ref{eq:expansion}), can be calculated using the well known Fourier method. The time dependent wave function of the charged particle is obtained using the calculated values of $c_n$'s and 1500 normalized stationary states of the pondermotive potential. The contribution of $c_n$'s in the time dependent wave function is shown in Fig. \ref{fig:cns}. It is observed that $c_n$'s for higher $n$ are extremely small compared to that for the lower $n$. Hence, the contribution of the higher stationary states in the time dependent wave function is negligible compared to the lower stationary states of ponderomotive potential.

\begin{figure}
    \centering
    \includegraphics[width=8.6cm]{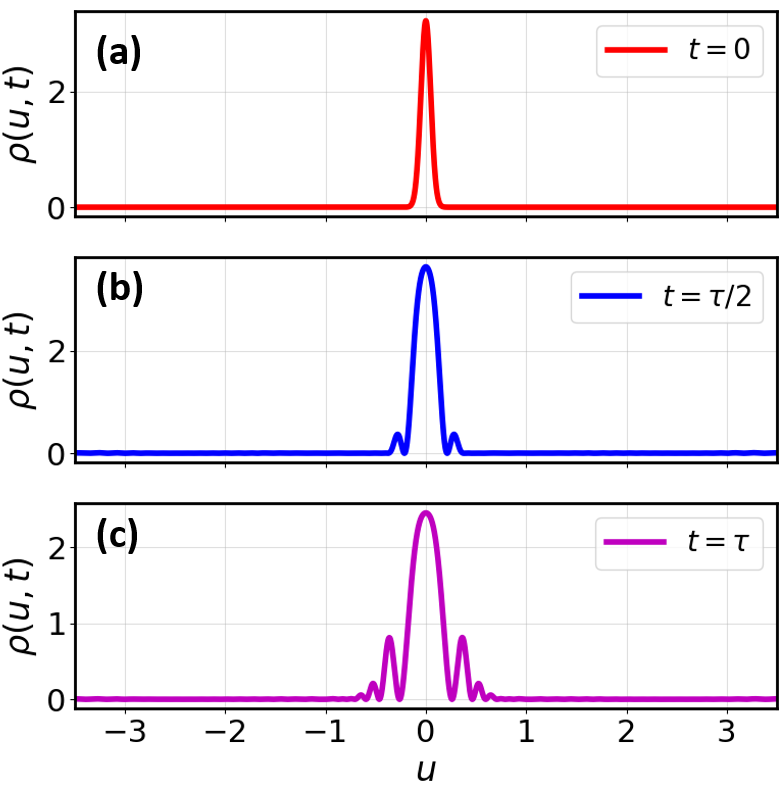}
    \caption{The probability density $\rho (u,t)$ of the particle at (a) $t=0$, (b) $t=\tau/2$, and (c) $t=\tau$. The interaction time $\tau$ ($\tau=d/v_e$, where $v_e=$ velocity of incident electrons in $z$ direction) is calculated as 0.8 ps.}
    \label{fig:diffraction}
\end{figure}

The time evolution of the probability density of the charged particles along the transverse direction is calculated by $\rho (u,t)=\psi_e(u,t) \psi_e^*(u,t)$. Figure \ref{fig:diffraction} shows the evolution of the particle wave function as they enter ($t=0$) and emerge out of the interaction region ($t=\tau$). At $t=0$, the probability density of the particle is given by  $\rho (u,0)$ (Fig. \ref{fig:diffraction}\textcolor{blue}{(a)}), which is due to the initial Gaussian wave function. Then $\rho$ evolves with time and at $t=\tau/2$, two more peaks on the either side of the central maxima ($m=0$) in $\rho(u,\tau/2)$ are observed, which correspond to the first order ($m=\pm 1$) diffraction peaks (Fig. \ref{fig:diffraction}\textcolor{blue}{(b)}). After the interaction ($t=\tau$), diffraction peaks for $m=0, \pm 1, \pm 2, \pm 3$ are observed (Fig. \ref{fig:diffraction}\textcolor{blue}{(c)}). The results of the time dependent probability density $\rho(u,t)$ during the interaction with optical crystal shows that particles get localized at equidistant positions along the transverse direction, which results in the diffraction pattern. The peaks in the diffraction pattern are equidistant, and the amplitudes are also according to the predicted value in Ref. \cite{Freimund8}.
\par 
%================================================================
% Diffraction of proton beams
%================================================================
\section{\label{sec:Parametric_study}Diffraction of proton beams} 
While until now the diffraction of electron and atomic sodium beams have been attempted, there have been no reports of studies of the KD effect for ion beams such as proton beams. Coherent (almost monoenergetic) proton beams can be obtained from a plasma confined in a multicusp \cite{Mathew,sanjeev2}. The effect of the strength and orientations of external perturbations such as electric and magnetic fields on the coherently split proton beams will be interesting to study experimentally. Through the use of an optical crystal, it may provide answers to questions such as whether a change in the orientation of the spin in a proton beam will have any effect on the diffraction pattern; this will belong to another area of physics that can be investigated \cite{Asma,Ahrens,Matthias}.

\begin{figure}[h]
    \centering
    \includegraphics[width=8.6cm]{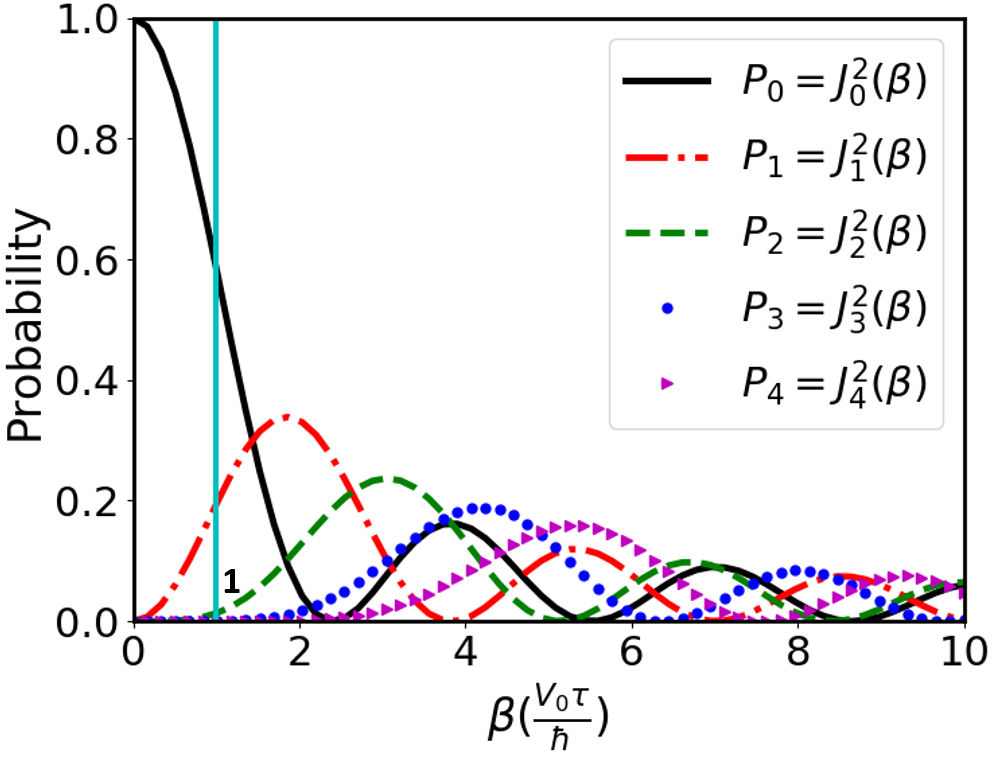}
    \caption{Probability distribution of the diffraction peaks with $\beta$ for first five orders ($m$ = 0, 1, 2, 3 and 4).}
    \label{fig:bessel_functions}
\end{figure} 

The probability to detect a particle in $m$th order diffraction peak is given by \cite{H_Batelaan},
\begin{equation}
P_m=J_m^2 \left( \frac{V_0 \tau}{\hbar}\right),
\end{equation}
where the argument of the Bessel function ($\beta=V_0 \tau/\hbar$), is proportional to the product of $V_0$ (strength of the Ponderomotive potential) and $\tau$ (interaction time). $\beta$ plays an important role in optimizing the experimental parameters such as energy ($E_i$) of the incident proton beam, intensity ($I$) and wavelength ($\lambda$) of the laser beam.

The probability distribution of the diffraction peaks for different orders as a function of $\beta$ is shown in Fig. \ref{fig:bessel_functions}. From figure \ref{fig:bessel_functions}, it is clear that to achieve appreciable detection probability of first few diffraction peaks, $\beta$ should be of the order of unity. 

\begin{figure}
    \centering
    \includegraphics[width=8.6cm]{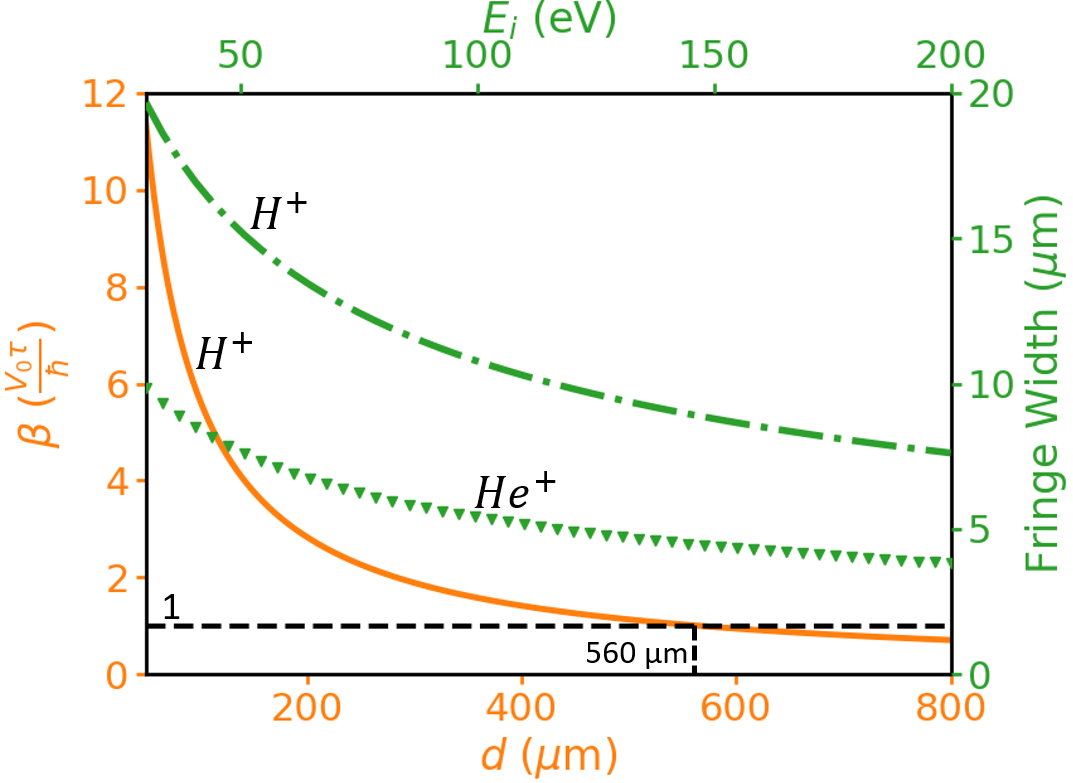}
    \caption{Dependence of $\beta$ on the laser beam waist ($d$) at the interaction region (orange-colored line) for $H^+$ having $E_i=50$ eV, and variation of fringe width of the diffraction pattern for $H^+$ (green-colored dotted line) and $He^+$ (green-colored triangles) beams with initial energy ($E_i$). Black-colored dotted line represents $\beta$ = 1. The curve for $\beta$ and the reference line ($\beta=1$), intersect at $d$ = 560 $\mu$m, therefore to observe the diffraction of proton beams with Nd: YAG laser ($\lambda=532$ nm, pulse width = 10 ns and $P=0.2$ J/pulse), the waist of the standing wave at the interaction region should be close to 560 $\mu$m.}
    \label{fig:beta_vs_beam_waist_proton}
\end{figure}

The interaction strength $\beta$ can be further expressed as,
\begin{equation}
\beta = \frac{e^2 \lambda^2 P}{2 \pi^3 \epsilon_0 c^3 \hbar d} \frac{1}{\sqrt{2mE_i}}.
\end{equation}
For fixed energy $E_i$ and other factors remaining constant, the variation of $\beta$ with $d$ for proton beams is shown in the Fig \ref{fig:beta_vs_beam_waist_proton}. It is clear that to obtain a detectable signal at different diffraction peaks, using Nd: YAG laser ($\lambda$ = 532 nm and $P$ = 0.2 J/ pulse), $d$ should be $\sim$560 $\mu$m, which is achievable using converging lens mirror system.

%@@@@@@

The fringe width ($W$) of the diffraction pattern will be decided by the following equation,
\begin{equation}
W=\frac{2 \lambda_{dB} D}{\lambda},
\end{equation}
where $\lambda_{dB}$ is the de Broglie wavelength of the matter-wave associated with the particles in the  incident beam which depends on the ion beam energy ($\lambda_{dB}=\hbar/\sqrt{2mE_i}$) and $D$ ($\sim$100 cm) is the distance between the interaction region and the detector. The variation of $W$ with $E_i$ for $H^+$ and $He^+$ beams is shown in Fig. \ref{fig:beta_vs_beam_waist_proton}. From figure \ref{fig:beta_vs_beam_waist_proton}, it is clear that lower the beam energy higher the fringe width ($W$),  which is in the range of few microns. Hence, in order to resolve the diffraction peaks, the detector should be moved in a step of size $\sim$1 $\mu$m or less. Also, it is observed that due to heavier mass of $He^+$, its $W$ is smaller than that of $H^+$ beams. \par 

From the above study, it is clear that to observe the KD effect for ion beams, it is better to use the lightest ion ($H^+$). The diffracted patterns can be resolved by moving the stage with the step size in submicron order, which is available in the current technology.

The energy of the ion beams $E_i$ as it emerges out of the plasma electrode aperture will be decided by the potential drop ($\sim$50 -70 V) in the plasma sheath. The plasma electrode is normally designed with a notch to provide a focusing effect. To further control the energy of the incoming ion beams, a puller electrode may be attached in front of the plasma electrode, and a well-collimated and focused ion beam with required beam energy can be obtained for the diffraction by an optical crystal. \par 

%================================================================
% Conclusions
%================================================================
\section{\label{sec:conclusions}Conclusions}
The stationary eigenstates and the eigenvalues of the ponderomotive potential are calculated by solving the time-independent Schrodinger's equation. In the numerical calculation, the Schrodinger's equation is solved by confining the ponderomotive potential in a one-dimensional infinite square well using a new method which can calculate the eigenvalues with much higher accuracy.  The calculated wave functions are symmetric and anti-symmetric, as expected from the nature of the ponderomotive potential. It is found that the eigenvalues of the ponderomotive potential are well fitted with the quadric polynomial of $n$, which can be explained by the mixing of the harmonic oscillator and the infinite square well. It is observed that the contribution of the square well in the eigenvalues is $\sim$0.59\% of the contribution of the harmonic oscillator. The dynamics of the charged particles in the nonlinear fields of optical crystal is studied to predict the diffraction of the incident beams. The time evolution of the charged particles is calculated by solving the time-dependent Schrodinger's equation. While solving the Schrodinger's equation, the initial state of the charged particle is expanded in the basis of actual stationary states of the ponderomotive potential. It is observed that the probability density of the particles gets localized along the direction of laser wave vectors, which results in the diffraction patterns. It is seen that the position of the diffraction patterns and the peak amplitudes are as expected from the theory available in the literature. To observe the diffraction of plasma-based proton beams, the diffraction phenomena for proton beams is investigated. It is found that the diffraction for plasma-based proton beams can be observed using the current technology, which can be interesting to fabricate interferometers of proton beams. The spin dynamics for proton beams in the presence of external electric and magnetic fields, and their effect on the diffraction pattern may be an interesting aspect of future experimental study. 
%================================================================
% Acknowledgements
%================================================================
%================================================================
% Reference 
%================================================================

\end{document}